\documentclass[10pt]{article}
\usepackage{graphicx}
\usepackage{amsfonts}
\usepackage{amssymb}

\newtheorem{theorem}{Theorem}
\newtheorem{definition}[theorem]{Definition}
\newtheorem{lemma}[theorem]{Lemma}
\newtheorem{notation}[theorem]{Notation}
\newtheorem{proposition}[theorem]{Proposition}
\newtheorem{remark}[theorem]{Remark}
\newenvironment{proof}[1][Proof]{\textbf{#1.} }{\ \rule{0.5em}{0.5em}}
\let\text=\mbox
\def\supp{\mathop\mathrm{supp}\nolimits}
\def\div{\mathop\mathrm{div}\nolimits}
\newcommand{\norm}[1]{\left\| #1 \right\|}
\newcommand{\abs}[1]{\left| #1 \right|}
\title{Bohmian trajectories and Klein's paradox}

\author{Gebhard Gr\"{u}bl, Raimund Moser and Klaus Rheinberger\\[10pt]Institut
f\"{u}r Theoretische Physik der Universit\"{a}t Innsbruck\\Technikerstr.
25, A-6020 Innsbruck\\[5pt]E-mail: gebhard.gruebl@uibk.ac.at}

\date{}

\begin{document}

\maketitle 

\begin{abstract}
We compute the Bohmian trajectories of the incoming scattering plane waves for
Klein's potential step in explicit form. For finite norm incoming scattering
solutions we derive their asymptotic space-time localization and we compute
some Bohmian trajectories numerically. The paradox, which appears in the
traditional treatments of the problem based on the outgoing scattering
asymptotics, is absent.
\end{abstract}


\section{Introduction}

Bohmian mechanics \cite{Bohm Hiley}\cite{HollandBuch} attempts to reconcile
quantum mechanics first with the notion of observation independent properties
of physical systems (``realism'') and second with strictly deterministic basic
laws connecting these properties (``determinism''). While in standard quantum
mechanics a system acquires a specific property only upon measurement by
random quantum jumps, within Bohmian mechanics each individual system has all
its possible properties completely specified independently of any measurement.
This is achieved by completing the quantum states through ``hidden
variables'', which determine all outcomes of experiments with individual systems.
Whether the laws,
which connect hidden variables, can be tested experimentally, is still under
debate. \cite{Neum}\cite{Marchildon} In any case the basic aim of Bohmian
mechanics has been appreciated by a growing community, \cite{Cushing} and
apart from all fundamental controversies, Bohmian trajectories undoubtedly
help to visualize the time evolution of wave functions.

While the main bulk of work within Bohmian mechanics has been devoted to the
Schr\"{o}dinger dynamics, there exist some results concerning relativistic
quantum mechanics. Among these is Holland's work on the Klein paradox.
\cite{HollandArtikel} In this work the Bohmian trajectories are computed,
which follow from the treatment of Klein's Paradox as it is given by Bjorken
and Drell. \cite{BjDr} Bjorken and Drell present a plane wave solution of the
Dirac equation for an electron, which is exposed to a sufficiently high, one
dimensional potential step. They talk about the various parts of their
solution as incoming, transmitted, and reflected waves. The paradox then
arises because the reflected probability current is greater than the incoming
current and the transition current is directed towards the potential step.
Holland \cite{HollandArtikel} determined the Bohmian trajectories associated
with the plane wave solution of Bjorken and Drell. He showed that they are
time like, future directed and do not intersect. Furthermore they do not begin
or end at finite time. Thus there is no indication for the production or
annihilation of electron-positron pairs in these trajectories. However, there
is something strange with them. They emerge from the region with nonzero
potential instead of moving into this region in the course of time. Furthermore
the trajectories of the physically interpretable finite norm wave packets have
not been considered at all in \cite{HollandArtikel}.

Already long ago Bongaarts and Ruijsenaars \cite{BongRuijAnnals} pointed out
the reason for the paradoxically directed transition current of Bjorken and
Drell's treatment: the chosen sign for the momentum parameter in the region
with nonzero potential. As a consequence, the plane waves of Bjorken and Drell
obey the boundary condition of an outgoing scattering solution instead of an
incoming one: A wave packet superposition of these waves, is localized within
the potential free half space and propagates away from the potential step at
large positive times. At large negative times the wave packet approaches the
potential step from both sides. Thus the surrounding talk of ``incoming'',
``transmitted'', and ``reflected'' waves, which was used by Bjorken and Drell,
is mathematically unjustified and misleading.

In contrast to these outgoing wave packets, a wave
packet superposition of the corresponding incoming plane wave solutions, is
localized within the potential free half space and propagates towards the
potential step at large negative times. At large positive times it moves away
from the potential step in both directions. Thus in this case the picture of
an incoming, transmitted, and reflected wave is based on mathematical facts.
Elaborate numerical studies of the evolution of Gaussian wave packets have
been made in order to confirm the completely different behaviour of the two
types of wave packets. \cite{Hörtn} As a result, the paradox of a negative
transition current, as formulated by Bjorken and Drell and investigated by
Holland, disappears, if the incoming scattering solutions are used in place
of the outgoing ones.

In this work we first discuss the Bohmian mechanics, associated with the incoming
scattering solutions of Klein's paradox, which initially move in the potential
free region towards the potential step. For the building block plane waves we
prove that all Bohmian trajectories move into the region with nonzero
potential. Second, we consider the finite norm wave packets. We derive their asymptotic
space-time localization properties and compute some trajectories numerically.
We find that at large negative times, all trajectories, which carry a
substantial part of the total norm, are located in the region with zero
potential and are directed towards the potential step, where some of them are
reflected and some are transmitted. At large positive times these trajectories
move away from the potential step. No indication for pair creation at the step
can be found. There remains, however, as will be shown here, a physically
questionable acceleration of slow packets upon passing the potential step.

\section{The 2d Dirac equation}

The set of space time points is assumed to be $M:=\mathbb{R}^{2}$. Let
$id_{M}=:(x^{0},x^{1})$ denote the standard chart of $M$. The associated
tangent frame is $\underline{\partial}:=(\partial_{0},\partial_{1})$. A
Minkowskian metric on $M$ is defined by $G_{p}(\partial_{\mu},\partial_{\nu
}):=\delta_{\mu}^{0}\delta_{\nu}^{0}-\delta_{\mu}^{1}\delta_{\nu}^{1}$ for
each $p\in M$. Let $\underline{e}=(e_{1},e_{2})$ be an orthonormal basis of a
$2$dimensional ($2$-d) $\mathbb{C}$-vector space $W$ with scalar product
$S:W\times W\rightarrow\mathbb{C}$. The linear mappings $\gamma^{\mu
}:W\rightarrow W$ are defined by $\gamma^{0}(e_{0})=e_{1}$, $\gamma^{0}%
(e_{1})=e_{0}$ and $\gamma^{1}(e_{0})=e_{1}$, $\gamma^{1}(e_{1})=-e_{0}.$ A
differentiable function $\psi:M\rightarrow W$ with
\begin{equation}
i\gamma^{\mu}\left(  \partial_{\mu}-i\frac{e}{\hbar c}A_{\mu}\right)
\psi=\kappa\psi\quad\text{(on }M\text{)}\label{Diracgl}
\end{equation}
is called a classical solution of the Dirac equation with continuous external
potential $A=A_{\mu}dx^{\mu} \equiv (A_{\mu})$  and Compton length $\kappa^{-1}\in
\mathbb{R}_{>0}$. The differential geometric notation is as in \cite{isham}.

Let $\psi^{1}$ and $\psi^{2}:M\rightarrow\mathbb{C}$ denote the (spinor-)
component functions of the function $\psi:M\rightarrow W$ with respect to the
basis $\underline{e}$, i.e. $\psi=:e_{1}\psi^{1}+e_{2}\psi^{2}$. The (Lorentz
invariant) indefinite inner product $L$ of the spinor space $W$ in terms of
its scalar product $S$ reads $L\left(  v,w\right)  :=S\left(  v,\gamma
^{0}w\right)  =v^{2\ast}w^{1}+v^{1\ast}w^{2}.$ The current vector field
$j_{\psi}:=j_{\psi}^{0}\partial_{0}+j_{\psi}^{1}\partial_{1} \equiv (j_{\psi}^\mu)$ of a
function $\psi:M\rightarrow W$ is defined by $j_{\psi}:=L(\psi,\gamma^{\mu}
\psi)\partial_{\mu}=\left(  \abs{\psi^{1}}^{2}+\abs{\psi^{2}}^{2}\right)
\partial_{0}+\left(  \abs{\psi^{1}}^{2}-\abs{\psi^{2}}^{2}\right)
\partial_{1}.$ Due to $G(j_{\psi},j_{\psi})=4\abs{\psi^{1}\psi^{2}}^{2}
\geq0$, the current is nowhere space-like. Where $j_{\psi}$ is nonzero, it is
future oriented. If $\psi$ is a classical solution of the Dirac equation
(\ref{Diracgl}), then $\div(j_{\psi})=0$.

For sufficiently regular potential, \cite{thaller} from $\div(j_{\psi})=0$ and
$j_{\psi}^{0}\geq0$ a probability structure for the spaces $\Sigma_{\tau
}:=\left\{  p\in M:x^{0}(p)=\tau\right\}  $ with $\tau\in\mathbb{R}$ can be
established as follows. The norm of the restriction $\psi_{\tau}$ of a
classical solution $\psi$ to $\Sigma_{\tau}$ for any $\tau\in\mathbb{R}$ is
defined by
\[
\norm{\psi_{\tau}} :=\left(  \int_{-\infty}^{\infty}j_{\psi}
^{0}(\tau,\xi)d\xi\right)  ^{\frac{1}{2}}.
\]
For solutions with $\norm{\psi_{0}}<\infty$ the equation $\norm{\psi_{\tau}} =
\norm{\psi_{0}}$ holds for any $\tau\in\mathbb{R}$.
Therefore the density
\[
\rho_{\psi,\tau}:=\frac{j_{\psi}^{0}(\tau,\xi)}{\norm{\psi_{0}}^{2}}\abs{d\xi}
\quad\text{with }\xi:= \left. x^{1}\right|_{\Sigma_{\tau}}
\]
is a probability density on $\Sigma_{\tau}$.

It has been suggested in sect. 12.2 of \cite{Bohm Hiley},
\cite{HollandBuch} that, due to $\div(j_{\psi})=0$, the density
$\rho_{\psi,\tau}$ is the transport of $\rho_{\psi,0}$ from
$\Sigma_{0}$ to $\Sigma_{\tau}$ along the flow lines of
$j_{\psi}$. The phenomenon may be visualised as the evolution of
the mass distribution of a cloud of dust along the individual
particle trajectories. In consequence, the set of flow lines of
$j_{\psi}$, the Bohmian trajectories, have been taken seriously as
the possible particle world lines, i.e. each (one particle) system
in the quantum state represented by $\psi$ is supposed to realise
one of the flow lines of $j_{\psi}$ in the course of time.

A general set of potentials and initial conditions (including
singular ones) seems to be unknown such that the global Bohmian
trajectories densely fibre $\supp(j_{\psi})$. (A global trajectory
is one, which extends both unboundedly into the past and into the
future. A dense fibration is such that the set $X\subset
\supp(j_{\psi})$ of points, which lie on a global Bohmian
trajectory, obeys $\rho_{\psi,\tau}(X\cap\Sigma_{\tau})=1$ for any
$\tau\in\mathbb{R}$.) The analogous problem in the Schr\"{o}dinger
case has been described by Berndl in \cite{Cushing} and resolved
in \cite{Berndl et al} for a wide class of potentials.

Lacking such general results for the Dirac equation we confine ourselves to
the very specific case of a discontinuous potential step. Let $0\leq
V\in\mathbb{R}$ and $\Theta:\mathbb{R}\rightarrow\mathbb{R}$ denote the step
function $\Theta(x\geq0):=1$ and $\Theta(x<0):=0$. Then the differential
$1$-form $A:=A_{0}dx^{0}$ with $-\frac{e}{\hbar c}A_{0}:=V\cdot\left(
\Theta\circ x^{1}\right)  $, defined on $U:=\left\{  p:x^{1}(p)\neq0\right\}
$, is introduced as an external potential into the restriction of
(\ref{Diracgl}) to the domain $U$. Only classical solutions of the restricted
equation with a continuous extension to $M$ are taken into consideration. This
yields the following system of partial differential equations for
differentiable component functions $\psi^{i}:U\rightarrow\mathbb{C}$ with
continuous extension to $M$.
\begin{equation}
i\partial_{0}\left(
\begin{array}[c]{r}
\psi^{1}\\
\psi^{2}
\end{array}
\right)  =\left(
\begin{array}[c]{cc}
-i\partial_{1}+V\Theta(x^{1}) & \kappa\\
\kappa &  i\partial_{1}+V\Theta(x^{1})
\end{array}
\right)  \left(
\begin{array}[c]{r}
\psi^{1}\\
\psi^{2}
\end{array}
\right)  \text{\quad(on }U\text{)}\label{DiracglU}
\end{equation}

For any such solution $\psi$, the current $j_{\psi}$ is continuous on $M$ and
differentiable on $U$. On $U$ the continuity equation $\div(j_{\psi})=0$ holds.

\section{Localization of free wave packets}

The finite norm solutions of (\ref{DiracglU}) with $V>0$ will be
constructed from the plane wave solutions of (\ref{Diracgl}) in
the case $A=0$. Thus this case is summarised first. See e.g.
\cite{BongRuijPoinc}.

\begin{notation}
Let $\overline{\omega}:\mathbb{R}\rightarrow\mathbb{R}_{>0}$ with
$\overline{\omega}(k):=\sqrt{\kappa^{2}+k^{2}}$ and $\Omega:\mathbb{R}
\rightarrow\mathbb{R}_{>0}$ with $\Omega(k):=\sqrt{\overline{\omega}(k)+k}$.
Then $u:\mathbb{R}\rightarrow W,v:\mathbb{R}\rightarrow W$ with
\[
u(k):=\underline{e}\cdot\left(
\begin{array}[c]{c}
\Omega(k)\\
\Omega(-k)
\end{array}
\right)  ,\qquad v(k):=\underline{e}\cdot\left(
\begin{array}[c]{c}
-\Omega(k)\\
\Omega(-k)
\end{array}
\right)  .
\]
\end{notation}

\begin{remark}
\label{innprod}For $k\neq0$ both $(u(k),u(-k))$ and $(v(k),v(-k))$ is a basis
of $W$. The following relations hold.
\[
\begin{array}{ll}
S(u(k),u(k)) = 2\overline{\omega}(k), &\qquad S(v(k),v(k))=2\overline{\omega}(k),\\
S(u(k),u(-k)) = 2\kappa, &\qquad S(v(k),v(-k))=2\kappa,\\
L(u(k),\gamma^{1}u(k)) = 2k, &\qquad L(v(k),\gamma^{1}v(k))=2k,\\
L(u(k),\gamma^{1}u(-k)) = 0, &\qquad L(v(k),\gamma^{1}v(-k))=0.
\end{array}
\]
\end{remark}

\begin{lemma}
Let $k\in\mathbb{R}\setminus0$ and let the function $f:\mathbb{R}\rightarrow
W$ be differentiable. Then a) and b) hold.

a) $\exp(-i\overline{\omega}(k)x^{0})f(x^{1}):M\rightarrow W$ solves the Dirac
equation (\ref{Diracgl}) with $A=0$ if and only if for some $\alpha,\beta
\in\mathbb{C}$
\[
f\left(  x^{1}\right)  =\alpha\exp(ikx^{1})u(k)+\beta\exp(-ikx^{1})u(-k).
\]

b) $\exp(i\overline{\omega}(k)x^{0})f\left(  x^{1}\right)  :M\rightarrow W$
solves the free Dirac equation (\ref{Diracgl}) with $A=0$ if and only if for
some $\alpha,\beta\in\mathbb{C}$
\[
f\left(  x^{1}\right)  =\alpha\exp(-ikx^{1})v(k)+\beta\exp(ikx^{1})v(-k).
\]
\end{lemma}

\begin{notation}
For $k\in\mathbb{R}$ we denote $\sqrt{2\pi}U_{k}:=\exp(-i(\overline{\omega
}(k)x^{0}-kx^{1})u(k),$ and $\sqrt{2\pi}V_{k}:=\exp(i(\overline{\omega
}(k)x^{0}-kx^{1})v(k).$
\end{notation}

The frequency of these plane wave solutions belongs to $\left(  -\infty
,-\kappa\right)  \cup\left(  \kappa,\infty\right)  $. To each frequency within
this range a $2$-dimensional subspace of single frequency solutions to
(\ref{Diracgl}) exists. The space of solutions with frequency $\overline
{\omega}(k)>\kappa$ is spanned by $(U_{k},U_{-k})$ and the space of solutions
with frequency $-\overline{\omega}(k)<-\kappa$ is spanned by $(V_{k},V_{-k})$.
Both $U_{k}$ and $V_{k}$ are constant along the space-like phase velocity
vector field $f:=\partial_{0}+\frac{\overline{\omega}(k)}{k}\partial_{1}$. The
current vector field, associated with both $U_{k}$ and $V_{k}$, is future
oriented, time like, and constant:
\[
j_{U_{k}}=j_{V_{k}}=\frac{1}{\pi}(\overline{\omega}(k)\partial_{0}
+k\partial_{1}).
\]

From the plane wave solutions $U_{k}$ and $V_{k}$\ of (\ref{Diracgl}) finite
norm wave packets are formed by superposition. Let $I\subset\mathbb{R}
$ be a compact interval and let the function $a:I\rightarrow
\mathbb{C}$ be continuous. Then the functions from $I\times M$ into $W$, with
either $(k,p)\mapsto a(k)U_{k}(p)$ or $(k,p)\mapsto a(k)V_{k}(p)$, first are
continuous and second have continuous partial derivatives with respect to
$x^{\mu}$. Therefore the functions from $M$ into $W$ with either $p\mapsto
\int\nolimits_{I}a(k)U_{k}(p)dk$ or $p\mapsto\int\nolimits_{I}a(k)V_{k}(p)dk$
are differentiable and the differentiation may be interchanged with the
integration and these functions are solutions of (\ref{Diracgl}) with $A=0$.

\begin{notation}
For a continuous function $a:I\rightarrow\mathbb{C}$, defined on a compact
real interval $I$, wave packet solutions $U\left[  a\right]  $ and $V\left[
a\right]  $ of (\ref{Diracgl}) are defined by
\[
U\left[  a\right]  :=\int\nolimits_{I}d\mu(k)a(k)U_{k}\text{ and }V\left[
a\right]  :=\int\nolimits_{I}d\mu(k)a(k)V_{k}\text{ with }d\mu(k):=\frac
{dk}{2\overline{\omega}(k)}.
\]
$U\left[  a\right]  $ is called positive frequency packet, $V\left[  a\right]
$ is called negative frequency packet.
\end{notation}

The movement of ``narrow'' wave packets $U\left[  a\right]  $ and $V\left[
a\right]  $ can be made plausible by replacing the function $\overline{\omega}
$ in the factor $\exp(-ix^{0}\overline{\omega})$ by its tangent approximation
at a point $k_{0}$ from the domain $I$. This yields $U\left[  a\right]  \simeq
P_{+}\left[  a\right]  U_{k_{0}}$ and $V\left[  a\right]  \simeq P_{-}\left[
a\right]  V_{k_{0}}$, with $v_{g}^{0}:=\frac{k_{0}}{\overline{\omega}(k_{0})}$
and
\[
P_{\pm}\left[  a\right]  :=\int_{I}d\mu(k)a(k)\exp\left[  \mp i(k-k_{0}%
)\cdot(v_{g}^{0}x^{0}-x^{1})\right]  .
\]
The functions $P_{\pm}\left[  a\right]  $, which modulate the plane waves
$U_{k_{0}}$ and $V_{k_{0}}$, are constant along the (future directed, time
like) group velocity vector field $g:=\partial_{0}+v_{g}^{0}\cdot\partial_{1}$
on $M$. Thus the sign of $k_{0}$ determines the direction of propagation of
$P_{\pm}\left[  a\right]  $ and in case of $k_{0}>0$ the approximations to
both $U\left[  a\right]  $ and $V\left[  a\right]  $ propagate towards growing
$x^{1}$.

A more conclusive derivation of the space-time localization of the wave
packets $U\left[  a\right]  $ and $V\left[  a\right]  $ follows from
proposition (3.1) of ref. \cite{Farina}. In the present case of one space
dimension this proposition reads as follows.

\begin{proposition}
\label{mosley}Let $\mathcal{F}:L^{2}(\mathbb{R})\rightarrow L^{2}(\mathbb{R})$
denote the Fourier transformation, formally given by
\[
\mathcal{F}\left(  f\right)  :k\mapsto\frac{1}{\sqrt{2\pi}}\int_{-\infty
}^{\infty}\exp(-ikx)f(x)dx.
\]
Let $\omega:\mathbb{R}\rightarrow\mathbb{R}$ be twice continuously
differentiable. The first derivative of $\omega$ is denoted as
$\omega ^{\prime}$. Define for any $t\in\mathbb{R}$ the unitary
time evolution operator $u_{t}:L^{2}(\mathbb{R})\rightarrow
L^{2}(\mathbb{R})$ through
$u_{t}(f):=\mathcal{F}^{-1}(\exp(-i\omega t)\mathcal{F}(f))$. Let
$f\in L^{2}(\mathbb{R})$ and $v_{1},v_{2}\in\mathbb{R}$ be such
that $v_{1} <\omega^{\prime}(k)<v_{2}$ for all
$k\in\supp\mathcal{F}(f)$. Then
\[
\lim_{t\rightarrow\infty}\int_{-\infty}^{tv_{1}}\abs{u_{t}(f)(x)}
^{2}dx=\lim_{t\rightarrow\infty}\int_{tv_{2}}^{\infty}\abs{u_{t}(f)(x)}
^{2}dx=0\text{.}
\]
\end{proposition}

\begin{remark}
Since $\norm{f}_{L^{2}}^{2}=\norm{u_{t}(f)}_{L^{2}
}^{2}$ for all $t\in\mathbb{R}$, the above statement is equivalent to
\[
\lim_{t\rightarrow\infty}\int_{tv_{1}}^{tv_{2}}\abs{u_{t}(f)(x)}
^{2}dx=\norm{f}_{L^{2}}^{2}.
\]
$u_{t}(f)$ localises for $t\rightarrow\infty$ within the interval
$t\cdot\left[  v_{1},v_{2}\right]  $.
\end{remark}

\begin{remark}
The localization of $u_{t}(f)$ for $t\rightarrow-\infty$ can be obtained from
the limit $t\rightarrow\infty$ of the evolution $\widetilde{u_{t}}:=u_{-t}$,
which has the frequency function $\widetilde{\omega}:=-\omega$. In this case
$-v_{2}<\widetilde{\omega}^{\prime}(k)<-v_{1}$. Thus
\[
\norm{f}_{L^{2}}^{2}=\lim_{t\rightarrow\infty}\int_{-tv_{2}
}^{-tv_{1}}\abs{\widetilde{u_{t}}(f)(x)}^{2}dx=\lim_{t\rightarrow-\infty
}\int_{tv_{2}}^{tv_{1}}\abs{u_{t}(f)(x)}^{2}dx
\]
follows. Both limits are covered by
\[
\norm{f}_{L^{2}}^{2}=\lim_{t\rightarrow\pm\infty}\int
_{t\cdot\left[  v_{1},v_{2}\right]  }\abs{u_{t}(f)(x)}^{2}dx\text{.}
\]
\end{remark}

An application of proposition (\ref{mosley}) to the component functions of
$U\left[  a\right]  $ and $V\left[  a\right]  $ yields the following
localization for $\rho_{\psi,\tau}$.

\begin{proposition}
\label{mosleydirac}$v_{g}:=\overline{\omega}^{\prime}:\mathbb{R}%
\rightarrow\mathbb{R}$, $k\mapsto\frac{k}{\overline{\omega}(k)}$ denotes the
group velocity of $\overline{\omega}$. The function $a:\left[  k_{1}
,k_{2}\right]  \rightarrow\mathbb{C}$ be continuous. The real numbers $v_{1}$
and $v_{2}$ are chosen such that $v_{1}<v_{g}(k_{1})$ and $v_{g}(k_{2})<v_{2}
$. For $\psi\in\left\{  U\left[  a\right]  ,V\left[  a\right]  \right\}  $
then
\[
\lim_{\tau\rightarrow\pm\infty}\int_{\tau\left[  v_{1},v_{2}\right]  }
\rho_{\psi,\tau}=1.
\]
\end{proposition}

\begin{proof}
We shall check first that the assumptions of proposition (\ref{mosley}) hold
for the component functions of our wave packets. Consider the positive
frequency case. Since the functions $\Omega$ and $\overline{\omega}$ are
continuous and since $\frac{1}{\overline{\omega}}$ is bounded, the function
\[
g^{i}:\mathbb{R}\rightarrow\mathbb{C},\quad k\mapsto\left\{
\begin{array}[c]{ll}
\frac{a(k)\Omega(-(-1)^{i}k)}{2\overline{\omega}(k)}&\qquad\text{for }
k_{1}<k<k_{2}\\
0&\qquad\text{otherwise}
\end{array}
\right.
\]
belongs to $L^{2}(\mathbb{R})$. Now $U\left[  a\right]  ^{i}$ is given by
$U\left[  a\right]  _{\tau}^{i}=\mathcal{F}^{-1}\left(  \exp(-i\overline
{\omega}\tau)g^{i}\right)  $. As $\overline{\omega}$ has continuous
derivatives of arbitrary order, the evolution is of the type of proposition
(\ref{mosley}). The second derivative $\overline{\omega}^{\prime\prime
}(k)=\frac{\kappa^{2}}{\overline{\omega}(k)^{3}}>0$ implies that the group
velocity function $k\mapsto\overline{\omega}^{\prime}(k)=\frac{k}
{\overline{\omega}^{\prime}(k)}$ is strictly increasing. From this one obtains
the bounds $v_{1}<\overline{\omega}^{\prime}(k)<v_{2}$ for any $k\in\left[
k_{1},k_{2}\right]  $. Thus proposition (\ref{mosley}) yields
\[
\norm{\left. U[a]^{i}\right|_{\Sigma_{\tau}}}_{L^{2}}
^{2}=\lim_{\tau\rightarrow\pm\infty}\int_{\tau\cdot\left[  v_{1},v_{2}\right]
}\abs{U[a]^{i}(\tau,x)}^{2}dx\text{.}
\]
From this and $\norm{U[a]_{\tau}}^{2}=\norm{U[a]_{\tau}^{1}}_{L^{2}}^{2}+
\norm{U[a]_{\tau}^{2}}_{L^{2}}^{2}$ the statement follows. The case of
negative frequency packets is analogous.
\end{proof}

\begin{remark}
Proposition (\ref{mosleydirac}) states that the probability density
$\rho_{\psi,\tau}$ is localized within the $x^{1}$ interval $\tau\cdot\left[
v_{1},v_{2}\right]  $ for $\tau\rightarrow\pm\infty$. In case of
$0<k_{1}<k_{2}$ it is localized in the half space $x^{1}<0$ for $\tau
\rightarrow-\infty$ and in the half space $x^{1}>0$ for $\tau\rightarrow
\infty$. The localization is right moving. In case of $k_{1}<k_{2}<0$ it is
localized in the half space $x^{1}>0$ for $\tau\rightarrow-\infty$ and in the
half space $x^{1}<0$ for $\tau\rightarrow\infty$. It is left moving.
\end{remark}

\section{Plane waves for $V>2\kappa$}

Let $\psi=\exp(-i\omega x^{0})f(x^{1})$ with $f:\mathbb{R}\rightarrow W$ be a
single frequency solution of (\ref{DiracglU}) with the frequency $\omega>\kappa$.
With some $\alpha,\beta\in\mathbb{C}$ the equation $\psi=\alpha U_{k}+\beta U_{-k}$
holds on $M_{-}:=\left\{  p\in M:x^{1}(p)<0\right\}$. Here $k>0$ is determined by
$\overline{\omega}(k)=\omega$. On $M_{+}:=\left\{
p\in M:x^{1}(p)>0\right\}  $ the function $\exp(iVx^{0})\psi$ equals a
single frequency solution of equation (\ref{Diracgl}) with $A=0$. Its frequency
reads $\omega^{\prime}:=\omega-V$.

Klein's phenomenon occurs for $\omega^{\prime}<-\kappa$. Thus we restrict our
discussion to the case $\kappa<\omega<V-\kappa$. This implies $V>2\kappa$. In
that case
\[
\psi=\exp(-iVx^{0})\left[  \gamma V_{q}+\delta V_{-q}\right]  \quad\text{(on
}M_{+}\text{)}%
\]
with $\gamma,\delta\in\mathbb{C}$ and with $q>0$ being determined by
$-\overline{\omega}(q)=\omega^{\prime}=\overline{\omega}(k)-V$. The constants
$\alpha,\beta,\gamma,\delta$ are restricted by the condition that $\psi$ is
continuous, which is equivalent to
\begin{equation}
\alpha U_{k}(0,0)+\beta U_{-k}(0,0)=\gamma V_{q}(0,0)+\delta V_{-q}
(0,0).\label{stetgkt}
\end{equation}
Since $U_{k}(0,0)$ and $U_{-k}(0,0)$ are linearly independent, this system of
linear equations for $\left(  \alpha,\beta,\gamma,\delta\right)  $ is of rank
$2$. Thus the space of single frequency solutions $\psi$ to (\ref{DiracglU}) is
$2$-dimensional. Within this space there are several $1$d subspaces of
particular physical importance.

One of these spaces comprises the single frequency solutions with
$\delta=0$. Its relevance emerges from the asymptotic behaviour in
time of the wave packets formed from these solutions. As will be
discussed in the next section, the $x^{0}=\tau$ restrictions of
such wave packets localise for $\tau \rightarrow-\infty$ within
the half line $x^{1}<0$. Thus these packets have a right moving
incoming asymptotics. Similarly, the $\alpha=0$ plane waves are
the building blocks, from which incoming left moving packets are
formed through superposition. Wave packets built from either
$\alpha=0$ or $\delta=0$ plane waves correspond to the
``incoming'' scattering solutions of the general quantum
scattering theory. \cite{BongRuijPoinc} They have a well defined
half space localization and direction of movement for
$\tau\rightarrow-\infty$. ``Outgoing'' scattering solutions
approaching a wave packet which, for $\tau\rightarrow\infty$,
moves out exclusively towards $x^{1}\rightarrow -\infty$, are
obtained from $\gamma=0 $. These are the solutions, on which refs.
\cite{BjDr}, \cite{HollandArtikel} base their discussion of
Klein's phenomenon. Finally, ``Outgoing'' solutions, which
approach a wave packet, that, for $\tau\rightarrow\infty$,
exclusively moves out towards $x^{1}\rightarrow\infty$, are
obtained from $\beta=0$.

Up to a constant factor the single frequency solutions with $\delta=0$ are given
by the following lemma, which follows from the continuity condition
(\ref{stetgkt}) with $\alpha=1$ and $\delta=0$.

\begin{lemma}
\label{einstrlsg}Let $V>2\kappa>0$ and $k>0$ be such that $\overline{\omega
}(k)<V-\kappa$ holds. Let the function $\psi:M\rightarrow W$ be
continuous and such that for some $q>0$ and some $r,t\in\mathbb{C}$
\[
\psi=\left\{
\begin{array}[c]{ll}
U_{k}+rU_{-k}&\text{\qquad on }M_{-}\\
\exp(-iVx^{0})tV_{q}&\text{\qquad on }M_{+}
\end{array}
\right.
\]
holds. Then $\psi$ is a solution of (\ref{DiracglU}) if and only if 1), 2), and 3) hold.

1) $q$ is the unique solution of $\overline{\omega}(k)+\overline{\omega}(q)=V$
in $\mathbb{R}_{>0},$

2) $r=r(k):=\frac{-2\kappa V}{V^{2}-(k-q)^{2}},$

3) $t=t(k):=-2\frac{k}{\kappa}\frac{\Omega(k)\Omega(-q)}{V+k-q}$.
\end{lemma}

\begin{notation}
The solution $\psi$ of (\ref{DiracglU}), which is given by lemma
(\ref{einstrlsg}) is denoted by $U_{k}^{in}$ in what follows.
\end{notation}

\begin{remark}
Observe that due to $V^{2}-(k-q)^{2}=2(\kappa^{2}+\overline{\omega
}(k)\overline{\omega}(q)+kq)>0$ and due to $V+k-q=\overline{\omega
}(k)+k+\overline{\omega}(q)-q>\overline{\omega}(q)-q>0$ the inequalities
$r(k)<0$ and $t(k)<0$ hold.
\end{remark}

\begin{remark}
\label{s}The wave number $q$ is given explicitly through $q=s(k)$, with the
differentiable function
\[
s:(0,\sqrt{V^{2}-2\kappa V})\rightarrow(0,\sqrt{V^{2}-2\kappa V}),\qquad
k\mapsto\sqrt{V^{2}-2V\overline{\omega}(k)+k^{2}}.
\]
The function $s$ is a monotonically decreasing bijection. It has the fixed
point $k_{0}=\frac{V}{2}\sqrt{1-\left(  \frac{2\kappa}{V}\right)  ^{2}}$. Thus
if $k<k_{0}$ then $q>k_{0}$ and if $k>k_{0}$ then $q<k_{0}$. These
inequalities will show up in the group velocity of narrow wave packets through
either an acceleration or a deceleration upon transition through the
potential's singularity at $x^{1}=0$.
\end{remark}

\begin{proposition}
For the current vector field $j_{U_{k}^{in}}$
\[
\pi j_{U_{k}^{in}}=t(k)^{2}\left[  \overline{\omega}(q)\partial_{0}
+q\partial_{1}\right]  +2\kappa r(k)\Theta(-x^{1})\left[  \cos(kx^{1}
)-1\right]  \partial_{0}
\]
holds. $j_{U_{k}^{in}}$ is differentiable and $\div(j_{U_{k}^{in}})=0$ everywhere on
$M$.
\end{proposition}

\begin{proof}
From remark (\ref{innprod}) one obtains by inserting the restrictions of
$U_{k}^{in}$ to $M_{\pm}$ into the current's definition the equations
\[
\pi j_{U_{k}^{in}}=\left\{
\begin{array}[c]{ll}
\left[  \overline{\omega}(k)\left(  1+r(k)^{2}\right)  +2\kappa r(k)\cos
(kx^{1})\right]  \partial_{0}+k\left(  1-r(k)^{2}\right)  \partial_{1}&\text{on }M_{-}\\
t(k)^{2}\left\{  \overline{\omega}(q)\partial_{0}+q\partial_{1}\right\}
&\text{on }M_{+}
\end{array}
\right.  .
\]
The continuity of $U_{k}^{in}$ implies the continuity of $j_{U_{k}^{in}}$,
which in turn is equivalent to the equations
\[
\begin{array}{l}
\overline{\omega}(k)\left(  1+r(k)^{2}\right)  +2\kappa r(k) =\overline
{\omega}(q)t(k)^{2},\\
k\left(  1-r(k)^{2}\right) =qt(k)^{2}.
\end{array}
\]
(They are easily checked by inserting the explicit expressions for $r(k)$ and
$t(k)$.) From this the formula for $j_{U_{k}^{in}}$ follows on $M$. Though
$U_{k}^{in}$ is not differentiable where $x^{1}=0$, its current field is
differentiable in every $p\in M$ because of $\left[  \cos(kx^{1})-1\right]
(p)=0$ and $\left(  \partial_{\mu}\left[  \cos(kx^{1})-1\right]  \right)
(p)=0$ for $x^{1}(p)=0$. Obviously, $\div(j_{U_{k}^{in}})=0$ holds on
$M_{+}\cup M_{-}$. Since $j_{U_{k}^{in}}$ is differentiable on $M$, it follows
that $\div(j_{U_{k}^{in}})=0$ on $M$.
\end{proof}

\begin{remark}
From the $j^{1}$-continuity condition $\left(  1-r(k)^{2}\right)
=\frac{qt(k)^{2}}{k}>0$ and from $r(k)<0$ the bounds $-1<r(k)<0$ follow.
\end{remark}

\begin{remark}
Since the Lie bracket $\left[  \partial_{0},j_{U_{k}^{in}}\right]  =0$, the
current vector field $j_{U_{k}^{in}}$ is $x^{0}$-shift invariant.
\end{remark}

\begin{definition}
Let $j$ be a differentiable vector field on $M.$ Let $\gamma_{p}:I\rightarrow
M$ obey the differential equation $\dot{\gamma}_{p}=j\circ\gamma_{p}$ and the
initial condition $\gamma(0)=p$. The open interval $I\subset\mathbb{R}$ is
assumed to be maximal. Since $j$ is differentiable, $\gamma_{p}$ is unique.
$\gamma_{p}$ is called the maximal integral curve of $j$ through $p\in M$ and the
set $\gamma_{p}(I)\subset M$ is called the orbit of $\gamma_{p}$. If $j$ is the
current vector field of a solution $\psi$ of the Dirac equation, then the
orbit of $\gamma_{p}$ is called the Bohmian trajectory of $\psi$ through $p$.
\end{definition}

\begin{remark}
$j_{U_{k}^{in}}$ is a bounded and everywhere future directed time
like vector field with positive, constant component
$j_{U_{k}^{in}}^{1}$. Thus along a Bohmian trajectory of
$U_{k}^{in}$ the function $x^{1}$ increases with $x^{0}$, i.e. the
trajectories move towards the potential step from $x^{1}<0$. To be
more specific: On $M_{+}$ the equation $j_{U_{k}^{in}
}=t(k)^{2}j_{U_{q}}$ holds. On $M_{-}$ the current
$j_{U_{k}^{in}}$ is the sum of $t(k)^{2}j_{U_{q}}$ and the non
constant $\partial_{0}$-directed, bounded vector field
$\frac{2\kappa}{\pi}r(k)\left[  \cos(kx^{1})-1\right]
\partial_{0}$. Due to $r(k)<0$ the additional term belongs to $R_{\geq0}
\cdot\partial_{0}$, which implies
\[
0<t(k)^{2}j_{U_{q}}^{0}\leq j_{U_{k}^{in}}^{0}\leq t(k)^{2}j_{U_{q}}^{0}
+\frac{4\kappa}{\pi}\abs{r(k)}\quad\text{on }M_{-}.
\]
From this and $j_{U_{k}^{in}}=t(k)^{2}j_{U_{q}}$ (on $M_{+}$) it follows that,
due to
\[
G(j_{U_{k}^{in}},j_{U_{k}^{in}})\geq t(k)^{2}G(j_{U_{q}},j_{U_{q}}%
)=t(k)^{2}\frac{\kappa^{2}}{\pi^{2}}>0,
\]
$j_{U_{k}^{in}}$ is globally time like and future directed. Thus
the current has no zeros and is nowhere light like.
\end{remark}

\begin{remark}
The velocity vector field $v_{U_{k}^{in}}:=\frac{j_{U_{k}^{in}}^{1}}
{j_{U_{k}^{in}}^{0}}\partial_{1}$ of $j_{U_{k}^{in}}$ relative to the inertial
frame $(\partial_{0},\partial_{1})$ is given by
\[
v_{U_{k}^{in}}=q\left[  \overline{\omega}(q)-\frac{2\kappa r(k)}{t(k)^{2}
}\Theta(-x^{1})\left(  1-\cos(kx^{1})\right)  \right]  ^{-1}\partial_{1}.
\]
On $M_{+}$ it is the constant field $\frac{q}{\overline{\omega}(q)}
\partial_{1}$ and on $M_{-}$ it oscillates between the positive bounds
\[
\frac{q}{\overline{\omega}(q)+\frac{4\kappa\abs{r(k)}}{t(k)^{2}}}
\partial_{1}\leq v_{U_{k}^{in}}\leq\frac{q}{\overline{\omega}(q)}\partial_{1}.
\]
From the bounds of $v_{U_{k}^{in}}$ it is obvious that a Bohmian trajectory of
$U_{k}^{in}$ cannot have a higher velocity within the range $M_{-}$ than it
has within $M_{+}$, where its velocity is less than $1$. An explicit formula
for the Bohmian trajectories of $U_{k}^{in}$ is given by the following proposition.
\end{remark}

\begin{proposition}
The Bohmian trajectory $\Gamma_{k,\tau}$ of $U_{k}^{in}$ through $p_{0}
:=(\tau,0)$ is the set of all $p\in M$, on which
\begin{equation}
x^{0}-\tau=\frac{\overline{\omega}(q)}{q}x^{1}-\frac{2\kappa r(k)}{kt(k)^{2}
}\theta(-x^{1})\left[  kx^{1}-\sin(kx^{1})\right]  \label{btrajebw}%
\end{equation}
holds. $\left\{  \Gamma_{k,\tau}\right\}  _{\tau\in\mathbb{R}}$ is a disjoint
covering of $M$ and $\Gamma_{k,\tau}\cap\left\{  p\in M:x^{1}(p)=0\right\}
=p_{0}$.
\end{proposition}

\begin{proof}
The current $j:=j_{U_{k}^{in}}=a\partial_{0}+b\partial_{1}+c\Theta
(-x^{1})(1-\cos(kx^{1}))\partial_{0}$ with $\pi a:=\overline{\omega
}(q)t(k)^{2},$ $\pi b:=qt(k)^{2},$ $\pi c:=-2\kappa r(k)$ is differentiable on
$M$. With $\gamma^{\mu}:=x^{\mu}\circ\gamma$ the differential equation
$\dot{\gamma}=j\circ\gamma$ decomposes into $\dot{\gamma}^{1}=b$ and
\[
\dot{\gamma}^{0}=\left\{
\begin{array}[c]{ll}
a&\text{ for }\gamma^{1}>0\\
a+c(1-\cos(k\gamma^{1})&\text{ for }\gamma^{1}<0
\end{array}
\right.  .
\]
The initial condition $\gamma^{1}(0)=0$ thus implies $\gamma^{1}
(\lambda)=b\lambda$ for all $\lambda\in\mathbb{R}$. Inserting this into the
equation for the component $\gamma^{0}$ one obtains from $\gamma^{0}(0)=\tau$
\[
\gamma^{0}(\lambda)=\left\{
\begin{array}[c]{ll}
a\lambda+\tau&\text{ for }\lambda>0\\
\left(  a+c\right)  \lambda-\frac{c}{kb}\sin(kb\lambda)+\tau&\text{ for
}\lambda<0
\end{array}
\right.  .
\]
Thus the maximal integral curve $\gamma_{p_{0}}$ of $j$ through $p_{0}$ reads
\[
\gamma_{p_{0}}:\mathbb{R}\rightarrow M,\quad\gamma_{p_{0}}(\lambda
)=(\tau,0)+\lambda\cdot\left(  a,b\right)  +\Theta(-\lambda)\left(
c\cdot(\lambda-\frac{\sin(kb\lambda)}{kb}),0\right)  .
\]
Its orbit is the set $\Gamma_{k,\tau}$ of points $p\in M$, on which
\[
x^{0}-\tau=\frac{a}{b}x^{1}+\frac{c}{kb}\theta(-x^{1})\left[  kx^{1}
-\sin(kx^{1})\right]
\]
holds. Obviously $\Gamma_{k,\tau}=(\tau,0)+\Gamma_{k,0}$ holds. As $\left. x^{0}
\right|_{\Gamma_{k,\tau}}$ is expressed in terms of $\left. x^{1}\right|_{\Gamma_{k,\tau}}
$, the trajectories $\Gamma_{k,\tau}$ and $\Gamma_{k,0}$ therefore do not
intersect for $\tau\neq0$. In particular, $\Gamma_{k,0}$ intersects $\left\{
p\in M:x^{1}(p)=0\right\}  $ only at the single point $(0,0)$. Because
$\left. x^{1}\right|_{\Gamma_{k,\tau}}\rightarrow\mathbb{R}$ is a bijection, the Bohmian
trajectory of $j$ through an arbitrary point $p$ is obtained by the proper
choice of $\tau$. Thus $\left\{  \Gamma_{k,\tau}\right\}  _{\tau\in\mathbb{R}
}$ is a disjoint covering, a fibration of $M$.
\end{proof}

Figure (1) shows the trajectory $\Gamma_{k,0}$ for $V=2.25\kappa$ and
$k=\frac{\kappa}{2}$ within the space time region where $-600<\kappa x^{0}
<200$ and $-50<\kappa x^{1}<50$.
\begin{figure}[ht]
\centering
\includegraphics[trim=0.000000in 0.000000in -0.307514in
0.000000in,height=6.3241cm,width=8.3911cm]{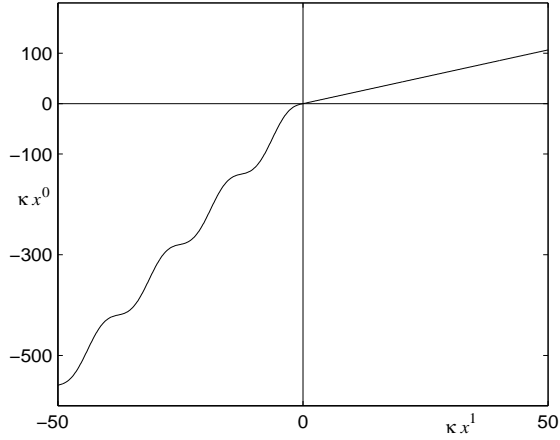}\\
\caption{$\Gamma_{k,0}$ for $k=\kappa/2$ and $V=2.25\kappa$}
\end{figure}

\section{Incoming localized solutions for $V>2\kappa$}

\begin{notation}
Let $a:I\rightarrow\mathbb{C}$ be continuous on a closed interval $I=\left[
k_{1},k_{2}\right]  \subset\mathbb{R}_{>0}$ of positive real numbers such that
on $I$ the inequality $\overline{\omega}<V-\kappa$ holds. The constants $V$
and $\kappa$ obey $V>2\kappa$. Then $U^{in}\left[  a\right]  :M\rightarrow W$
denotes the wave packet $U^{in}\left[  a\right]  :=\int\nolimits_{I}
d\mu(k)a(k)U_{k}^{in}$.
\end{notation}

\begin{proposition}
\label{Kleinpaket}$U^{in}\left[  a\right]  $ is continuous on $M$ and
differentiable on $U$. It is a solution of equation (\ref{DiracglU}). Let
$a_{r}:\left[  -k_{2},-k_{1}\right]  \rightarrow\mathbb{C}$ and $a_{t}:\left[
s(k_{2}),s(k_{1})\right]  \rightarrow\mathbb{C}$ be defined by $a_{r}
(k):=r(-k)a(-k)$, $a_{t}(q):=\frac{q}{k}t(k)a(k)$ with $k=\sqrt{V^{2}
-2V\overline{\omega}(q)+q^{2}}$. Then the equations $U^{in}\left[  a\right]
=U\left[  a\right]  +U\left[  a_{r}\right]$ on $M_{-}$ and $U^{in}\left[
a\right]  =\exp(-iVx^{0})V\left[  a_{t}\right]$ on $M_{+}$ hold.
\end{proposition}
 
\begin{proof}
From $U_{k}^{in}=U_{k}+r(k)U_{-k}$ (on $M_{-}$) there follows on $M_{-}$
\[
\int\nolimits_{k_{1}}^{k_{2}}d\mu(k)a(k)U_{k}^{in}=\int\nolimits_{k_{1}
}^{k_{2}}d\mu(k)a(k)U_{k}+\int\nolimits_{-k_{2}}^{-k_{1}}d\mu
(k)a(-k)r(-k)U_{k}.
\]
This proves $U^{in}\left[  a\right]  =U\left[  a\right]  +U\left[
a_{r}\right]  $ on $M_{-}$.

Similarly, with $q=s(k):=\sqrt{V^{2}-2V\overline{\omega}(k)+k^{2}}$ there
follows on $M_{+}$:
\[
\int\nolimits_{k_{1}}^{k_{2}}d\mu(k)a(k)U_{k}^{in}=\exp(-iVx^{0}
)\int\nolimits_{k_{1}}^{k_{2}}d\mu(k)a(k)t(k)V_{s(k)}.
\]
The function $s$ is defined implicitly by $\overline{\omega}+\overline{\omega
}\circ s=V$. This implies first $s=s^{-1}$ and second $d\overline{\omega
}=-d\left(  \overline{\omega}\circ s\right)  $. Substitution of the
integration variable $k$ by $q=s(k)$ then yields
\begin{eqnarray*}
\int\nolimits_{k_{1}}^{k_{2}}\frac{dk}{2\overline{\omega}(k)}a(k)t(k)V_{s(k)}
&=&\int\nolimits_{k_{1}}^{k_{2}}\frac{d\overline{\omega}(k)}{2k} a(k)t(k)V_{s(k)}\\
&=&-\int\nolimits_{s(k_{1})}^{s(k_{2})}\frac{d\overline{\omega}(q)}{2q}\frac
{q}{s(q)}a(s(q))t(s(q))V_{q}\\
&=&\int\nolimits_{s(k_{2})}^{s(k_{1})} d\mu(q)\frac{q}{s(q)}a(s(q))t(s(q))V_{q}.
\end{eqnarray*}
This proves $U^{in}\left[  a\right]  =\exp(-iVx^{0})V\left[
a_{t}\right]  $ on $M_{+}$. The statements about the continuity
and differentiability follow from the continuity of $a_{r}$ and
$a_{t}$ through application of elementary analysis theorems.
\end{proof}

\begin{remark}
The localization of $U^{in}\left[  a\right]  $ for $\tau\rightarrow\pm\infty$
is immediate from proposition (\ref{Kleinpaket}). Since $a_{r}$ has its domain
within $\mathbb{R}_{<0}$ and $a$ and $a_{t}$ have their domains within
$\mathbb{R}_{>0}$, the wave packet $U\left[  a_{r}\right]  $ is left moving,
while $U\left[  a\right]  $ and $U\left[  a_{t}\right]  $ are right moving. At
large negative times $U^{in}\left[  a\right]  $ approximates $U\left[
a\right]  $, which moves in through $M_{-}$ towards $x^{1}=0$. At large
positive times $U^{in}\left[  a\right]  $ approximates $U\left[  a_{t}\right]
+U\left[  a_{r}\right]  $. The transmitted wave $U\left[  a_{t}\right]  $
moves away from $x^{1}=0$ through $M_{+}$ and the reflected one, $U\left[
a_{r}\right]  $, moves away from $x^{1}=0$ through $M_{-}$.
\end{remark}

Rigourous analysis of whether the Bohmian trajectories of $\left.
U^{in}\left[ a\right] \right|_{M_{-}}$ may be connected with the
trajectories of $\left.U^{in}\left[  a\right] \right|_{M_{+}}$
across $x^{1}=0$ in a unique way is left to further investigation.
As an indication that this should be possible, we present some
trajectories, computed numerically.

For $V=4\kappa$ we choose for several values of the constants $\Delta>0$ and
$K>0$ the Fourier amplitude
\[
a:\left[  k_{1},k_{2}\right]  \rightarrow\mathbb{R},\quad a(k):=\exp\left(
\frac{-(k-K)^{2}}{\Delta^{2}}\right)
\]
with $k_{1}=K-2\Delta$, $k_{2}=K+2\Delta$. The domain of the function $s$ of
remark (\ref{s}) is $0<k<\sqrt{8}\kappa$ and its fixed point is $k_{0}
=\sqrt{3}\kappa$. Clearly, the constants $K$ and $\Delta$ have to be chosen
such that $\left[  k_{1},k_{2}\right]  $ is contained in the domain of $s$.
First the wave packet $U^{in}\left[  a\right]  $ is computed numerically in
that space-time region, where it hits the potential step. Second from the
associated current field some Bohmian trajectories are computed by numerical
integration. The starting points are chosen within $\Sigma_{\tau}$ around the
center of localization of $\rho_{U^{in}\left[  a\right]  ,\tau}$. The initial
time $\tau$ is such that the main bulk of the probability distribution
$\rho_{U^{in}\left[  a\right]  ,\tau}$ has not yet arrived at the potential step.

Figure (2) shows some trajectories for $\Delta=0.1\kappa$ and $K=0.3\kappa$
within the space-time region where $-180<\kappa x^{0}<200$ and $-100<\kappa x^{1}
<300$. Since the wave numbers $k$ from the domain of $a$ obey $k<k_{0}$
the domain of $a$ is mapped into $k>k_{0}$ by $s$ such that the transmitted
packet is faster than the incoming one. This might be considered as what is
left over from Klein's paradox.%
\begin{figure}[h]
\centering
\includegraphics[ trim=0.378876in 0.152205in 0.000000in
0.000000in,height=7.697cm,width=10.7327cm]{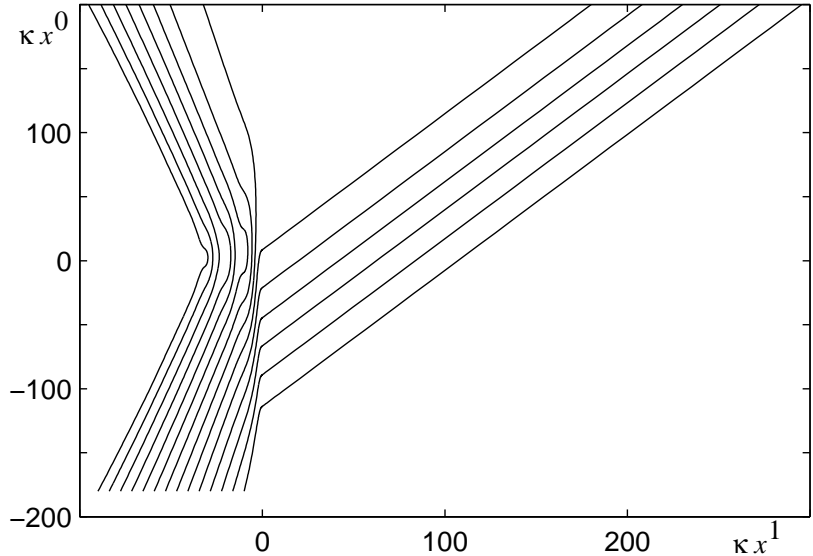}\\ 
\caption{trajectories for $\Delta=0.1\kappa$ and
$K=0.3\kappa$}
\end{figure}

Figure (3) shows some trajectories for $\Delta=0.05\kappa$ and $K=2.7\kappa$
within the space-time region where $-100<\kappa x^{0}<160$ and $-300<\kappa x^{1}
<200$. Since the wave numbers $k$ from the domain of $a$ obey $k>k_{0}$
the transmitted packet is slower than the incoming one. Many more pictures for
related situations are contained in ref. \cite{Moser}.
\begin{figure}[h]
\centering
\includegraphics[trim=0.312407in 0.166846in 0.000000in
-0.109807in,height=7.2752cm,width=9.9881cm]{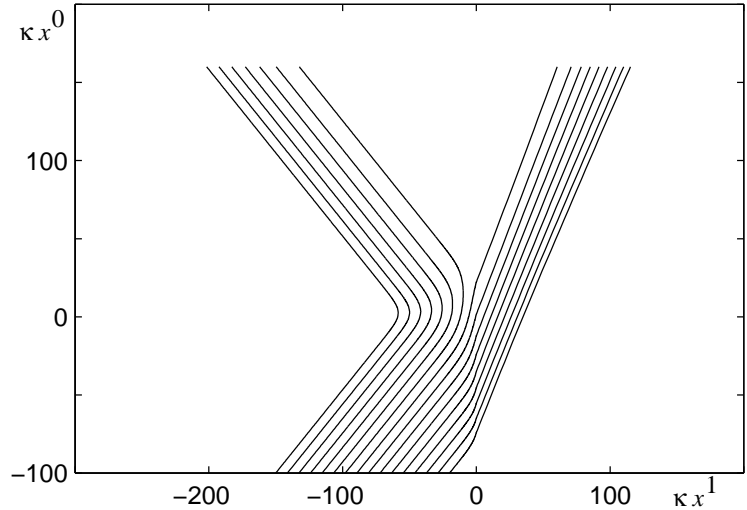}\\
\caption{trajectories for $\Delta=0.05\kappa$ and $K=2.7\kappa$}
\end{figure}

\section*{Acknowledgments} 
The authors are indebted to Dr. H. G. Embacher for \LaTeX\ support.

\end{document}